\documentclass[proof]{pasj01}
\SetRunningHead{T. Magara}{Evolution and dynamics of a solar active prominence}

\usepackage{times}
\usepackage{epsfig}

\begin{document}

\title{EVOLUTION AND DYNAMICS OF A SOLAR ACTIVE PROMINENCE}
\author{T. \textsc{Magara}}
\affil{Dept. of Astronomy and Space Science, School of Space Research,
Kyung Hee University,
1732 Deogyeong-daero, Giheung-gu,
Yongin, Gyeonggi-do, 446-701,
Republic of Korea}
\email{magara@khu.ac.kr}


\maketitle

\begin{abstract}
The life of a solar active prominence, one of the most remarkable objects on the Sun, is full of dynamics; after first appearing on the Sun the prominence continuously evolves with various internal motions and eventually produces a global eruption toward the interplanetary space. Here we report that the whole life of an active prominence is successfully reproduced by performing as long-term a magnetohydrodynamic simulation of a magnetized prominence plasma as was ever done. The simulation reveals underlying dynamic processes that give rise to observed properties of an active prominence: invisible subsurface flows self-consistently produce the cancellation of magnetic flux observed at the photosphere, while observed and somewhat counterintuitive strong upflows are driven against gravity by enhanced gas pressure gradient force along a magnetic field line locally standing vertical. The most highlighted dynamic event, transition into an eruptive phase, occurs as a natural consequence of the self-consistent evolution of a prominence plasma interacting with a magnetic field, which is obtained by seamlessly reproducing dynamic processes involved in the formation and eruption of an active prominence. {\bf A manuscript with high-resolution figures is found at http://web.khu.ac.kr/$\sim$magara/index.html. Here `$\sim$' is a tilde.}  
\end{abstract}

\section{Introduction}
Solar prominences are, just as their name tells, one of the prominent objects forming in the solar corona (Tandberg-Hanssen 1995). They are also called filaments when observed as a dark thin and long object on a solar disk. Prominences have various sizes with different appearances; their physical nature has been investigated in detail via theoretical modeling and multi-wavelength observations, which has revealed that a magnetic field frozen in a prominence plasma plays the role of a skeleton in determining the structural and evolutionary properties of a prominence (Kippenhahn \& Schl{\"u}ter 1957; Kuperus \& Raadu 1974; Schmieder, Malherbe, Mein \& Tandberg-Hanssen 1984; Martin, Livi \& Wang 1985; van Ballegooijen \& Martens 1989; Choe \& Lee 1992; Low 1996; Aulanier \& Demoulin 1998; Martin 1998; DeVore \& Antiochos 2000; Martens \& Zwaan 2001; Chae, et al. 2001; Pevtsov, Balasubramaniam \& Rogers 2003; Lites 2005; Mackay \& van Ballegooijen 2005; Magara 2007; Okamoto et al. 2008; Berger et al. 2011; Priest 2014 and references therein).

Prominences are classified into {\it quiescent prominences} and {\it active prominences}; the former are found outside solar active regions and size and lifetime are larger and longer than the latter which form in active regions and frequently produce explosive phenomena such as solar flares. This suggests that the formation processes of these two types of prominences may be different (Priest 2014). Here we report that dynamic processes giving rise to observed properties of an active prominence are seamlessly reproduced by performing as long-term a magnetohydrodynamic simulation of a magnetized prominence plasma as was ever done. The simulation demonstrates how these dynamic processes are self-consistently driven and maintained.

\section{Simulation setup}
The simulation was performed in Cartesian coordinates $(x, y, z)$ with the $z$-axis directed upward. $z < 0$, $z=0$ and $z > 0$ correspond to the subphotosphere, the solar surface and the solar atmosphere extending above the surface, respectively. The simulation box is $(-200, -200, -10) < (x, y, z) < (200, 200, 190)$ which is discretized into non-uniform grid cells whose size is $(\Delta x, \Delta y, \Delta z) = (0.2, 0.2, 0.2)$ for $(-20,-20,-10) < (x, y,z) < (20,20,20)$ and it gradually increases up to $(4, 4, 4)$ as $|x|, |y|$ and $z$ increase. The total number of cells is $N_x \times N_y \times N_z = 383 \times 383 \times 242$. The normalization units of the simulation are given by 2$H_{ph}$ (length), $c_{ph}$ (velocity), $\rho_{ph}$ (gas density), $\rho_{ph}c_{ph}^2$ (gas pressure), $T_{ph}$ (temperature) and $(\rho_{ph}c_{ph}^2)^{1/2}$ (magnetic field), where $H_{ph}$, $c_{ph}$, $\rho_{ph}$ and $T_{ph}$ represent pressure scale height, sound speed, gas density and temperature at the photosphere. As the initial state of the simulation we adopt the same state as described in Magara (2012): a cylindrical magnetic flux tube of a Gold-Hoyle profile with a radius $r_f=2$ and twist parameter $b=1$ (Gold \& Hoyle 1960) is placed horizontally in the subphotosphere (the axis of the flux tube is set 4 below the $y$-axis). The flux tube keeps mechanical equilibrium with a background atmosphere stratified under the solar surface gravity ${g}_{\odot} =$ 274 m/s$^2$ and a prescribed temperature profile based on a solar atmospheric model (Vernazza, Avrett \& Loeser 1981). Boundary conditions and wave damping zones placed near all the boundaries are the same as the ones used in Magara (2012). The simulation is based on viscous magnetohydrodynamic equations with the Reynolds number Re $ \sim 2.5 \times 10^4$ where a length scale is given by the typical height of an active prominence $l \sim 7,000$ km and a velocity scale is the free-fall velocity $v_f \equiv (2 {g}_{\odot}l)^{1/2} \sim 20$ km/s. The simulation started with the rise of the flux tube via the magnetic buoyancy (Parker 1955), which forms three emerging lobes of a magnetic field at intervals of 40, as shown by Figure 1a. Emergence of a twisted flux tube (Emonet \& Moreno-Insertis 1998; Fan 2001; Magara \& Longcope 2003; Archontis et al. 2004; Manchester IV et al. 2004; Galsgaard et al. 2005; Cheung \& Isobe 2014 and references therein) is, observationally as well as theoretically, suggested to be one of the possible formation processes of an active prominence (Rust \& Kumar 1994; Low 1996; Magara 2007; Okamoto et al. 2008). The emerging portion of the flux tube that enters a low atmosphere ($0 \le z \le 4$) is subject to the Newton's cooling characterized by the relaxation time $\tau_r \sim 1$ s (Stix 1991). 

\section{Result and discussion}
Figures 1b-d display snapshots of the middle emerging lobe taken at different evolutionary phases, which are viewed from the subphotosphere to see how a flow and magnetic field evolve below the surface when a prominence forms above the surface. These figures show a transition process from a divergent flow at an early phase during which the area of photospheric magnetic flux increases (figure 1b) to a convergent flow at a late phase when positive and negative photospheric flux approach each other (figure 1d), known as {\it flux cancellation} (Martin, Livi \& Wang 1985).

While flux cancellation is operating at the surface, a magnetic field emerging into the solar atmosphere develops a structure on the Sun, which is composed of the core field lines that form the main body of a prominence lying in the horizontal position and the envelop field lines overlying the main body (figure 2a). The gas density in the main body is about $10^{5}$ or $10^{6}$ times smaller than that in the photosphere, which is consistent with observations of active prominences (Priest 2014). Figures 2a-d show the evolution of selected field lines whose colors indicate the value of gas density. As the main body gradually changes shape from a horizontal shape into an $\Omega$ shape, the gas density in the main body continues to decrease. This is because a prominence plasma easily falls down along an $\Omega$-shaped field line, which in turn enhances the magnetic buoyancy of $\Omega$-part of the field line, facilitating the transition of field-line shape. The simulation demonstrates a positive feedback between the draining of mass and the transition of field-line shape, suggesting that it is inevitable for the prominence to proceed to an eruptive phase (figure 2d).

Strong upflows of several tens of km/s exist in a prominence (Priest 2014) and their driving mechanism has been an issue to be clarified because the speed is comparable to the typical free-fall speed mentioned in section 2. In this simulation there are found locations of such strong upflows, one of which is indicated by the white box in figure 2d. Figure 3 presents a close-up view of this upflow location, where a field line is plotted to see the configuration of a magnetic field and distribution of gas pressure along the field line. The color gradient of the field line represents the variation of gas pressure, which explains that the upflows are driven by strong gas pressure gradient force along the field line locally standing vertical.

The global eruption of an active prominence, which is an event at the final stage of evolution, is often associated with other explosive phenomena such as a solar flare and/or coronal mass ejection (Shibata \& Magara 2011). They are transient but intense solar phenomena which potentially make a severe impact on the terrestrial environment. In order to understand a mechanism for producing these harmful phenomena and ultimately predict them, we have to clarify the dynamic properties of an active prominence evolving to global eruption. Seamlessly reproducing dynamic processes involved in the formation and eruption of an active prominence presented here gives an important step toward the comprehensive understanding and prediction of solar explosive phenomena.

%%the dynamic nature of an active prominence is one of the fundamental issues to be clarified. The self-consistent evolution of an active prominence reproduced by the present simulation gives important insights into a transition mechanism for dynamical states of a continuously evolving prominence plasma toward global eruption.

\bigskip
The author wishes to thank the Kyung Hee University for general support of this work. He also appreciates useful comments on the paper by S. Solanki. Visualization of simulation data was done using VAPOR (Clyne et al. 2007). This work was financially supported by the Basic Science Research Program (2013R1A1A2058705, PI: T. Magara) through the National Research Foundation of Korea (NRF) provided by the Ministry of Education, Science and Technology as well as the BK21 program (20132015) through the NRF.

\newpage

\begin{figure}
  \begin{center}
\includegraphics*[width=15cm]{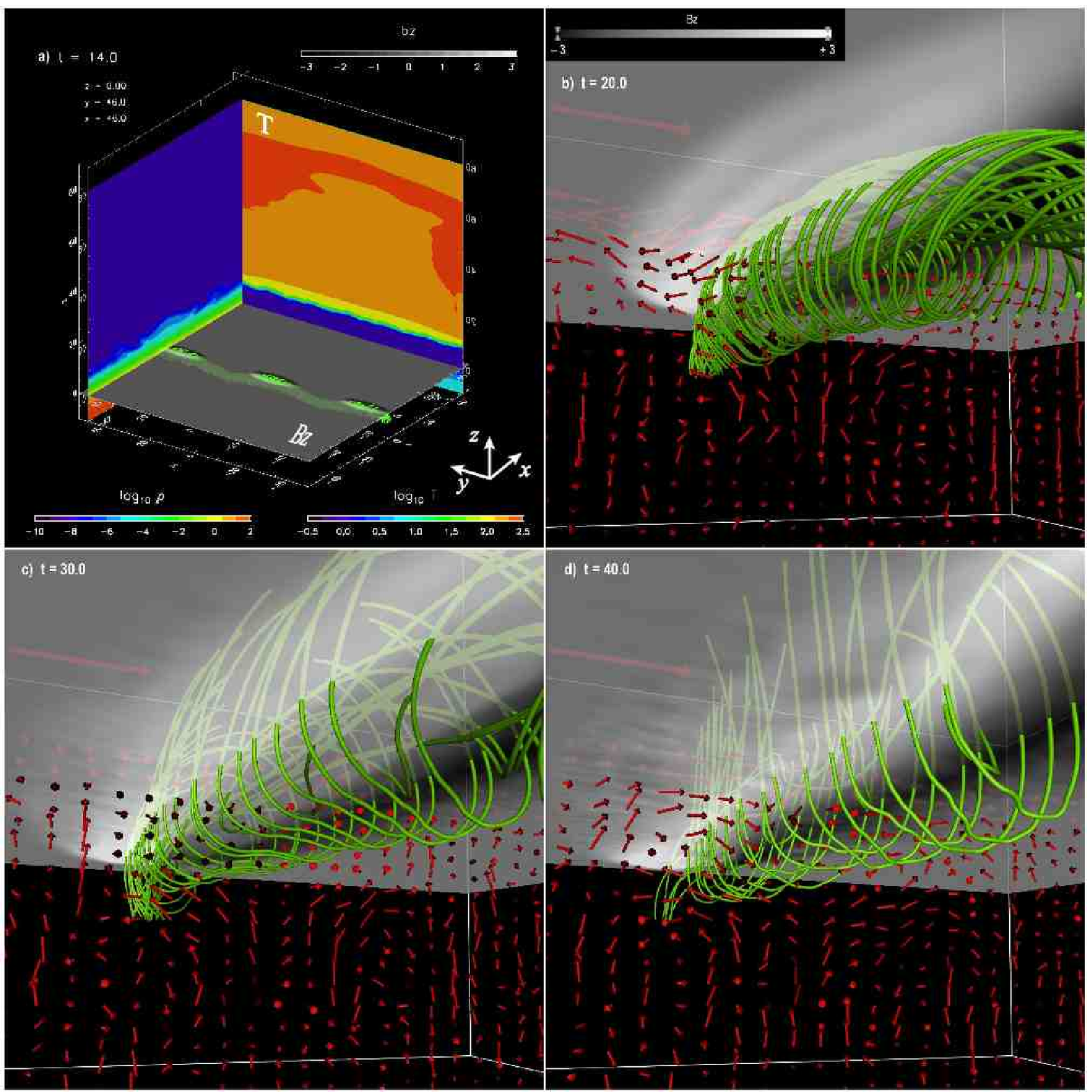}
  \end{center}
\caption{(a): Snapshot of a local simulation domain $(-50, -50, -10) < (x, y, z) < (50, 50, 90)$ taken at $t=14$ is presented. The green lines represent twisted field lines which form three emerging lobes while the gray-scale map at $z=0$ gives the surface distribution of vertical magnetic flux in linear scale. Two color maps placed at $x=46$ and $y=46$ show the distributions of temperature and gas density in logarithmic scale. Gas density, temperature and magnetic field are normalized by photospheric density $\rho_{ph}$, photospheric temperature $T_{ph}$ and $(\rho_{ph}c_{ph}^2)^{1/2}$ where $c_{ph}$ is photospheric sound speed. (b)-(d): Snapshots of the middle emerging lobe taken at $t=20$ (Figure 1b), 30 (Figure 1c) and 40 (Figure 1d) show the evolution of selected field lines drawn in green. The gray-scale map placed at $z=0$ shows the surface distribution of vertical magnetic flux in linear scale. The red arrows represent flow velocity field at $y=0$. A unit vector scaled to $c_{ph}$ is given at top-left corner of each figure. \label{fig1}}
\end{figure}

\begin{figure}
  \begin{center}
\includegraphics*[width=15cm]{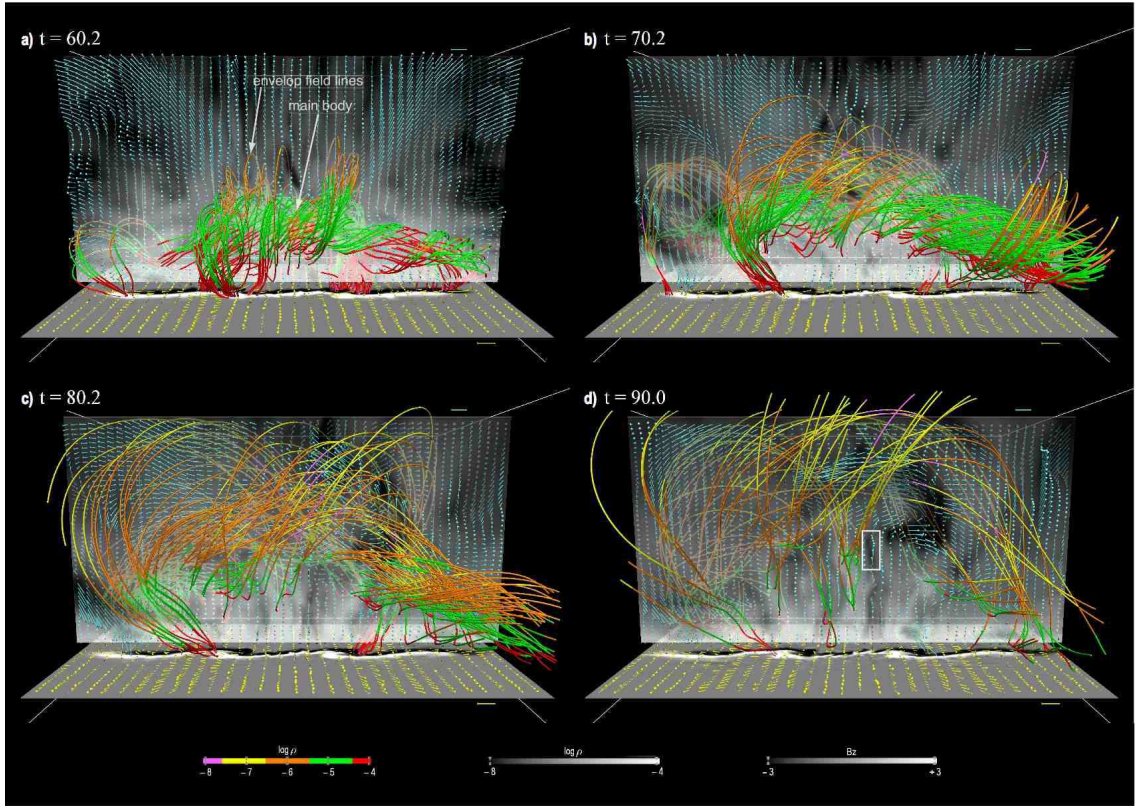}
  \end{center}
\caption{(a): Snapshot of a local simulation domain $(-80, -80, -10) < (x, y, z) < (80, 80, 90)$ taken at $t=60.2$ is presented. The colored lines represent selected field lines; colors indicate the value of gas density in logarithmic scale. The horizontal gray-scale map at $z=0$ shows the surface distribution of vertical magnetic flux in linear scale. The vertical gray-scale map at $x=0$ shows the distribution of gas density in logarithmic scale. The yellow arrows represent flow velocity field at $z=0$ and a unit vector scaled to $c_{ph}$ is given at bottom-right corner. The blue arrows represent flow velocity field at $x=0$ and a unit vector scaled to 10 $c_{ph}$ is given at top-right corner. (b)-(d): Snapshots of the same selected field lines as those in Figure 2a are presented, except for $t=70.2$ (Figure 2b), $t=80.2$ (Figure 2c) and $t=90$ (Figure 2d). The white box in Figure 2d shows a location of strong upflows, the close-up view of which is presented in Figure 3. \label{fig2}}
\end{figure}

\begin{figure}
  \begin{center}
\includegraphics*[width=15cm]{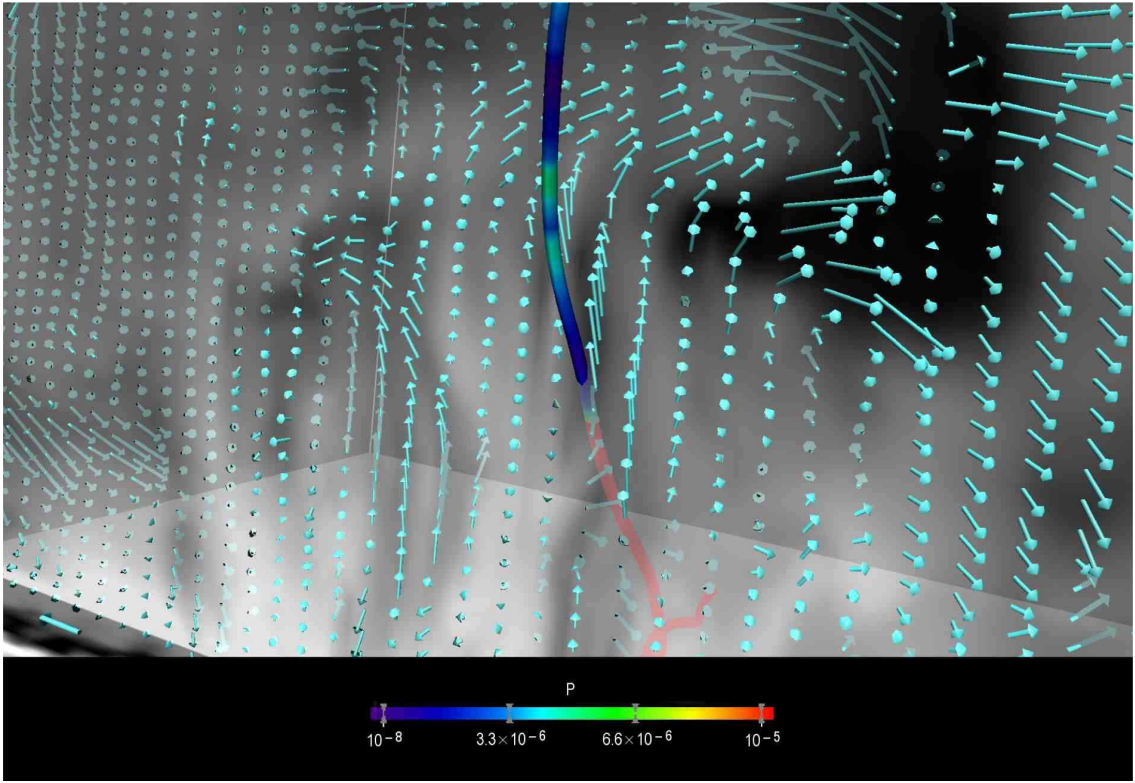}
 \end{center}
\caption{Close-up view of the upflow location indicated by the white box in Figure 2d is presented. The color gradient of the displayed field line shows the variation of gas pressure in linear scale, normalized by photospheric gas pressure $P_{ph}=\rho_{ph}c_{ph}^2 \gamma^{-1}$ where $\gamma$ is the specific heat ratio. A unit vector scaled to 10 $c_{ph}$ is given at bottom-left corner.
\label{fig3}}
\end{figure}


\clearpage


\begin{thebibliography}{}

\bibitem[Archontis et 
al.(2004)]{2004A&A...426.1047A} Archontis, V., Moreno-Insertis, F., Galsgaard, K., Hood, A., \& O'Shea, E.\ 2004, \aap, 426, 1047 

\bibitem{1998A&A...329.1125A} Aulanier, G. \& Demoulin, P.
\ 1998, \aap, 329, 1125

\bibitem[Berger et al.(2011)]{2011Natur.472..197B} Berger, T., Testa, P., 
Hillier, A., et al.\ 2011, \nat, 472, 197 

\bibitem{2001ApJ...548..497C} Chae, J. et al.
\ 2001, \apj, 548, 497

\bibitem[Cheung 
\& Isobe(2014)]{2014LRSP...11....3C} Cheung, M.~C.~M., \& Isobe, H.\ 2014, Living Reviews in Solar Physics, 11, 3

\bibitem[Choe 
\& Lee(1992)]{1992SoPh..138..291C} Choe, G.~S., \& Lee, L.~C.\ 1992, \solphys, 138, 291 

\bibitem{2007NJPh....9..301C} Clyne, J., Mininni, P., 
Norton, A., \& Rast, M.\ 
{\it New Journal of Physics}, 9, 301 (2007)

\bibitem{2000ApJ...539..954D} DeVore, C.~R. \& Antiochos, S.~K.
\ 2000, \apj, 539, 954

\bibitem[Emonet 
\& Moreno-Insertis(1998)]{1998ApJ...492..804E} Emonet, T., \& Moreno-Insertis, F.\ 1998, \apj, 492, 804 

\bibitem[Fan(2001)]{2001ApJ...554L.111F} Fan, Y.\ 2001, \apjl, 554, L111 

\bibitem[Galsgaard et al.(2005)]{2005ApJ...618L.153G} Galsgaard, K., 
Moreno-Insertis, F., Archontis, V., \& Hood, A.\ 2005, \apjl, 618, L153 

\bibitem[Gold 
\& Hoyle(1960)]{1960MNRAS.120...89G} Gold, T., \& Hoyle, F.\ 1960, \mnras, 120, 89 

\bibitem{1957ZA.....43...36K} Kippenhahn, R. \& Schl{\"u}ter, A.
\ 1957, {\it Zeitschrift fur Astrophysik}, 43, 36

\bibitem{1974A&A....31..189K} Kuperus, M. \& Raadu, M.~A.
\ 1974, \aap, 31, 189 

\bibitem[Lites(2005)]{2005ApJ...622.1275L} Lites, B.~W.\ 2005, \apj, 622, 
1275 

\bibitem[Low(1996)]{1996SoPh..167..217L} Low, B.~C.\ 1996, \solphys, 167, 
217 

\bibitem[Mackay 
\& van Ballegooijen(2005)]{2005ApJ...621L..77M} Mackay, D.~H., \& van Ballegooijen, A.~A.\ 2005, \apjl, 621, L77 

\bibitem[Magara 
\& Longcope(2003)]{2003ApJ...586..630M} Magara, T., \& Longcope, D.~W.\ 2003, \apj, 586, 630 

\bibitem[Magara(2007)]{2007PASJ...59L..51M} Magara, T.\ 2007, \pasj, 59, 
L51

\bibitem[Manchester et al.(2004)]{2004ApJ...610..588M} Manchester, W., IV, 
Gombosi, T., DeZeeuw, D., \& Fan, Y.\ 2004, \apj, 610, 588 

\bibitem[Martens 
\& Zwaan(2001)]{2001ApJ...558..872M} Martens, P.~C., \& Zwaan, C.\ 2001, \apj, 558, 872 

\bibitem[Martin et al.(1985)]{1985AuJPh..38..929M} Martin, S.~F., Livi, 
S.~H.~B., \& Wang, J.\ 1985, Australian Journal of Physics, 38, 929 

\bibitem{1998SoPh..182..107M} Martin, S.~F.
\ 1998, Sol. Phys., 182, 107


\bibitem[Okamoto et al.(2008)]{2008ApJ...673L.215O} Okamoto, T.~J., 
Tsuneta, S., Lites, B.~W., et al.\ 2008, \apjl, 673, L215

\bibitem{parker55} Parker, E. N.
\ 1955, \apj, 121, 491

\bibitem{2003ApJ...595..500P} Pevtsov, A.~A., Balasubramaniam, K.~S., \& Rogers, J.~W.
\ 2003, \apj, 595, 500

\bibitem[Priest(2014)]{2014masu.book.....P} Priest, E.\ 2014, 
Magnetohydrodynamics of the Sun, by Eric Priest, Cambridge, UK: Cambridge 
University Press, 2014,  

\bibitem{1994SoPh..155...69R} Rust, D.~M. \& Kumar, A.
\ 1994, Sol. Phys., 155, 69

\bibitem[Schmieder et 
al.(1984)]{1984A&A...136...81S} Schmieder, B., Malherbe, J.~M., Mein, P., \& Tandberg-Hanssen, E.\ 1984, \aap, 136, 81 


\bibitem{2006SoPh..238..245S} Schmieder, B., Aulanier, G., Mein, P., \& L{\'o}pez Ariste, A.
\ 2006, Sol. Phys., 238, 245

\bibitem[Shibata 
\& Magara(2011)]{2011LRSP....8....6S} Shibata, K., \& Magara, T.\ 2011, Living Reviews in Solar Physics, 8, 6 

\bibitem[Stix(1991)]{1991sun..book.....S} Stix, M.\ 1991, The Sun.~ An 
Introduction, XIII, 390 pp.~192 figs..~Springer-Verlag Berlin Heidelberg 
New York.~ Also Astronomy and Astrophysics Library,

\bibitem[Tandberg-Hanssen(1995)]{1995ASSL..199.....T} Tandberg-Hanssen, E.\ 
1995, Astrophysics and Space Science Library, 199,  

\bibitem[van Ballegooijen 
\& Martens(1989)]{1989ApJ...343..971V} van Ballegooijen, A.~A., \& Martens, P.~C.~H.\ 1989, \apj, 343, 971 

\bibitem[Vernazza et al.(1981)]{1981ApJS...45..635V} Vernazza, J.~E., 
Avrett, E.~H., \& Loeser, R.\ 1981, \apjs, 45, 635 

\end{thebibliography}
\end{document}